\documentclass[aip,rsi,reprint,superscriptaddress,floatfix]{revtex4-1}
\usepackage{amsfonts}
\usepackage{amsmath}
\usepackage{amssymb}
\usepackage{graphicx}%
\usepackage[breaklinks]{hyperref}
\providecommand{\U}[1]{\protect \rule{.1in}{.1in}}
\newlength {\defaultparindent }
\newenvironment {annotation Text}{}{}
\newcommand\includegraphicss[1]
  {\resizebox{8.5cm}{!}{\includegraphics[clip=true]{#1}}}
\newcommand \beginfigure{\begin{figure}[p]}

\begin{document}
\title{GINA -- A Polarized Neutron Reflectometer at the Budapest Neutron Centre}
\author{L.~Botty\'{a}n}
\email{bottyan.laszlo@wigner.mta.hu}
\affiliation{Wigner RCP, RMKI, P.O.B. 49, H-1525 Budapest, Hungary}
\author{D.~G.~Merkel}
\affiliation{Wigner RCP, RMKI, P.O.B. 49, H-1525 Budapest, Hungary}
\author{B.~Nagy}
\affiliation{Wigner RCP, RMKI, P.O.B. 49, H-1525 Budapest, Hungary}
\author{J.~F\"{u}zi}
\affiliation{Wigner RCP, SZFKI, P.O.B. 49, H-1525 Budapest, Hungary}
\affiliation{University of P\'{e}cs, PMMIK, P.O.B. 219, H-7602 P\'{e}cs, Hungary}
\author{Sz.~Sajti}
\affiliation{Wigner RCP, RMKI, P.O.B. 49, H-1525 Budapest, Hungary}
\author{L.~De\'{a}k}
\affiliation{Wigner RCP, RMKI, P.O.B. 49, H-1525 Budapest, Hungary}
\author{G.~Endr\H{o}czi}
\affiliation{Wigner RCP, RMKI, P.O.B. 49, H-1525 Budapest, Hungary}
\author{A.~V.~Petrenko}
\affiliation{Frank Laboratory of Neutron Physics, JINR, Dubna, Russia}
\author{J.~Major}
\affiliation{Wigner RCP, RMKI, P.O.B. 49, H-1525 Budapest, Hungary}
\affiliation{Max-Planck-Institut f\"{u}r Intelligente Systeme (formerly Max-Planck-Institut
f\"{u}r Metallforschung), Heisenbergstr. 3, D-70569 Stuttgart, Germany}
\date{\today }

\begin{abstract}
The setup, capabilities and operation parameters of the neutron reflectometer
GINA, the recently installed \textquotedblleft Grazing Incidence Neutron
Apparatus\textquotedblright \ at the Budapest Neutron Centre, are introduced.
GINA, a dance-floor-type, constant-energy, angle-dispersive reflectometer is
equipped with a 2D position-sensitive detector to study specular and
off-specular scattering. Wavelength options between 3.2 and
$5.7~\mathrm{\mathring{A}}$ are available for unpolarized and polarized
neutrons. Spin polarization and analysis are achieved by magnetized
transmission supermirrors and radio-frequency adiabatic spin flippers. As a
result of vertical focusing by a five-element pyrolytic graphite
monochromator, the reflected intensity from a $20\times20~\mathrm{mm}^{2}$
sample has been doubled. GINA is dedicated to studies of magnetic films and
heterostructures, but unpolarized options for non-magnetic films, membranes,
and other surfaces are also provided. Shortly after its startup, reflectivity
values as low as $3\times10^{-5}$ have been measured by the instrument. The
instrument capabilities are demonstrated by a non-polarized and a polarized
reflectivity experiment on a Si wafer and on a magnetic film of $\left[
^{62}\mathrm{Ni}/^{\mathrm{nat}}\mathrm{Ni}\right]  _{5}$ isotope-periodic
layer composition. The facility is now open for the international user
community. Its further development is underway establishing new sample
environment options and spin analysis of off-specularly scattered radiation as
well as further decreasing the background.

\end{abstract}
\keywords{}
\pacs{}
\maketitle
\preprint{HEP/123-qed}

%\tableofcontents

\section{Introduction\label{Introduction}}

The ever increasing need for product advancement and miniaturization keeps
membranes, thin film assemblies, magnetic and non-magnetic multilayer and
patterned heterostructures in the limelight of materials science and
technological development. In recent years more and more complex methods,
instrumental and evaluation types have emerged to meet the new challenges. Due
to the matching wavelength range of cold neutrons and their extreme
sensitivity to the interface structure and to the internal magnetic fields,
neutron reflectometry (NR) is a rather capable non-destructive method to
investigate such nanostructures. Reflected intensity measured as a function of
the momentum transfer $Q_{z}$, perpendicular to the sample surface provides
information on the scattering length density (SLD) depth profile. Normalized
reflectivities are usually recorded from unity in the total reflection regime
to $Q_{z}$ values of about $0.2~\mathrm{\mathring{A}}^{-1}$, where it drops by
five to six orders of magnitude. Layer thicknesses appear as regular features
in the reflected intensity and may be modeled using optical formalisms.
\cite{refe1,refe2}
Similar but spin dependent formalisms apply for polarized neutron
reflectometry (PNR) in magnetic studies. \cite{refe3,refe4,refe5} 
By fitting the appropriate model
to the measured intensities as a function of $Q_{z}$, one can extract layer
thickness, interfacial roughness, and depth-dependent SLD, as well as
magnetization profile. By measuring scattered intensity as a function of an
in-plane wave vector component $Q_{x}$, one can, in addition, characterize
lateral structures. \cite{refe6} In the lateral direction, the interface may be rough on
different length scales or may display a defined periodicity, resulting in
diffuse scattering or Bragg reflections in $Q_{x}$ scans, respectively.
Polarized neutron reflectometry (PNR) not only provides an isotope-selective
atomic density depth profile (as well in the case of deeply buried layers)
with a spatial resolution of a few nanometers, \cite{refe7,refe8} but it is also a highly
sensitive magnetometry to determine the vector properties of the
magnetization. The prototype polarized neutron reflectometer was developed and
built at Argonne in the 1980s. \cite{refe9,refe10} The increased interest in magnetic thin film
analytical instruments triggered by the discovery of the giant
magnetoresistance and related phenomena \cite{refe11,refe12} resulted in a boom of PNR studies 
\cite{refe13,refe14,refe15} as well as the construction of a number of new neutron reflectometers with
polarization option at neutron sources across the globe. 

The Budapest Research Reactor (BRR) of the Budapest Neutron Centre 
(BNC), Hungary operates some 15 instruments on thermal channels and cold 
guides. \cite{refe17} BRR is a tank type reactor, moderated and cooled 
by light water. The core flux is 
$2.1\times10^{14}\mathrm{n}/\mathrm{cm}^{2}\mathrm{s}$. 
The \textquotedblleft
Grazing Incidence Neutron Apparatus\textquotedblright \ (GINA), 
a polarized-beam neutron reflectometer has recently been 
installed \cite{refe16} on the cold guide  $10/3$ of BRR. Here we report 
on the design, the construction and the operation parameters of GINA. 
Although over 30 neutron reflectometers offer polarization option 
worldwide, \footnote{The actual list with links to the neutron 
reflection facilities can be found at the page of A.R. Rennie:\\ 
\protect\url{http://material.fysik.uu.se/Group_members/adrian/reflect.htm}} 
the overall available beam time on polarized reflectometers 
is generally overbooked, thus a new instrument and its performance may 
interest the growing user community. 

\section{GENERAL OVERVIEW OF THE GINA INSTRUMENT}

\begin{figure}[t]
%\resizebox{.99\columnwidth}{!}{\includegraphics{Fig2.eps}}
\includegraphicss {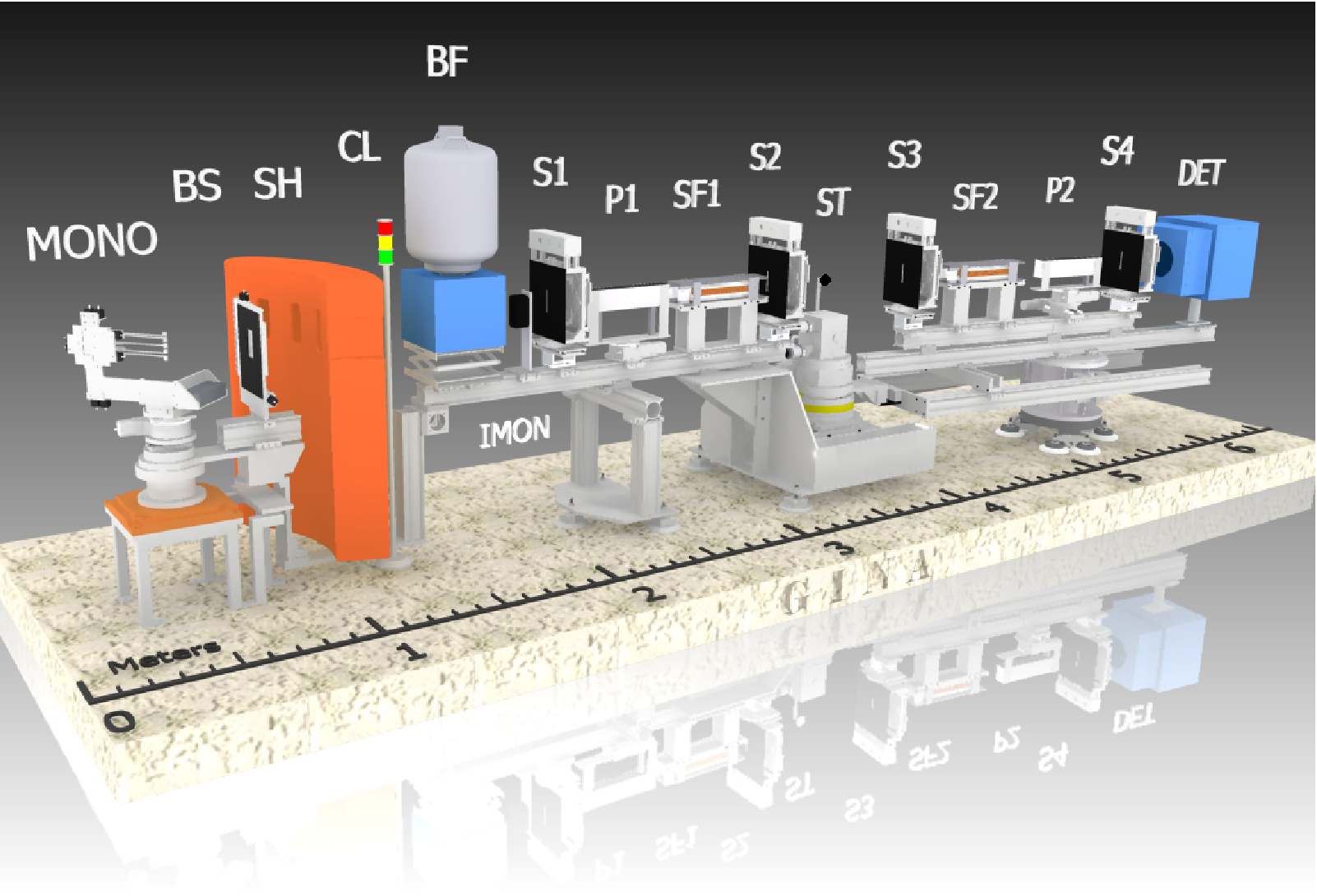}
 \caption{The layout of the GINA neutron reflectometer. The monochromator
assembly (MONO) is mounted behind the concrete shielding (SH) on a turntable
connected to the optical bench (B1) supporting the beam shutter (BS),
(monitored by the semaphore control light (CL)), the intensity monitor
detector (IMON), the cryostat of the beryllium filter (BF), the slit S1, the
magnetic supermirror polarizer (P1), the adiabatic RF spin flipper (SF1) and
the slit (S2). The bench is fixed to the central goniometer tower (ST), the
components of which define the position and orientation of the sample surface
relative to the beam and supports the components of the sample environment.
The optical bench (B2) is connected to the turn-table underneath the central
goniometer tower and it supports the slit (S3), spin flipper (SF2), the
magnetic supermirror spin analyzer (P2), and slit (S4) in front of the
detector unit (DET).
 }
 \label{Fig2}
\end{figure}

The GINA neutron reflectometer is a constant-energy angle-dispersive
instrument with a horizontal scattering plane. \cite{refe18} The monochromator assembly,
which is mounted in a gap of the cold neutron guide 10/3, selects the
wavelength of the monochromatic beam within the range of
$3.2~\mathrm{\mathring{A}}-5.7~\mathrm{\mathring{A}}$ and can focus the beam
in the vertical plane. In order to produce a polarized neutron beam, a
magnetized polarizing supermirror (PSM) in transmission geometry and an
adiabatic radio frequency (RF) spin flipper \cite{refe19,refe20} are used. The beam scattered by
the sample undergoes spin analysis by an identical setup of a spin flipper and
a spin filter, and finally it is detected by a two-dimensional position
sensitive detector. To reduce the background, the detector is encased in a
B$_{2}$O$_{3}$-mixed polyethylene shielding and four motorized slits
are provided in the beam path to discard the undesired neutrons, including
those scattered from the device components. The complete instrument setup is
shown in Fig.~\ref{Fig2}. The neutron-optical devices of the reflectometer are
mounted on two X95 optical benches 
\footnote{Newport Co., Irvine, CA, USA, \protect\url{www.newport.com}} 
%\cite{refe21} 
to provide accurate definition of the beam
height and a heavy-load support for various additional elements.
  \noindent
%\begin{figure}[p]

The optical bench B1 (cf. Fig.~\ref{Fig2}) defines the horizontal optical axis of the
reflectometer and supports the beam shutter, the intensity monitor detector,
the cryostat of the beryllium filter BF, the slit S1, the polarizer P1, the
spin flipper SF1 and the slit S2. Bypassing the radiation shielding by a
U-shape construction, the bench B1 is connected to the turntable of the
monochromator, a heavy-load goniometer with vertical axis. The axis of the
turntable coincides with that of the monochromator. The angle of the optical
axis relative to the guide and consequently the wavelength can be changed by
manually rotating the entire GINA setup around the turntable while air pads
are activated and the entire setup floats over the marble floor. The allowed
wavelengths are restricted at present to 3.2, 3.9, 4.6, 5.2 and
$5.7~\mathrm{\mathring{A}}$ by the respective channels through the cylindrical
concrete shielding around the monochromator unit.

The downstream end of the optical bench B1 is fixed to the central goniometer
tower ST, the components of which define the position and orientation of the
sample surface relative to the beam and supports the various sample
environment components (electromagnet, cryostat, etc.). The X95 optical bench
B2 -- the 2$\theta$-arm of the reflectometer -- is connected to the sample
turntable underneath the central goniometer tower and it supports the slit S3,
the spin flipper SF2, the spin analyzer P2, and the slit S4 in front of the
detector unit along with its electronics and dedicated control PC. The
encoder-controlled precise motion around the turntable is performed by a
rubber-coated motorized wheel under the respective while the air pads are
pressurized (cf. Fig.~\ref{Fig2}).

\section{WAVELENGTH SELECTION AND FOCUSING}

The monochromator is located 17.8 meters downstream from the cold source. The
curved Ni/Ti supermirror neutron guide between the cold source and the GINA
monochromator has a horizontal radius of $340~\mathrm{m}$, a height of
$100~\mathrm{mm}$ and a width of $25~\mathrm{mm}$. The inner, outer, top and
bottom mirrors are of $m=2,3,2$ and 2, respectively. 

\begin{figure}[t]
%\resizebox{.45\columnwidth}{!}{\includegraphics{Fig3.eps}}
\includegraphicss {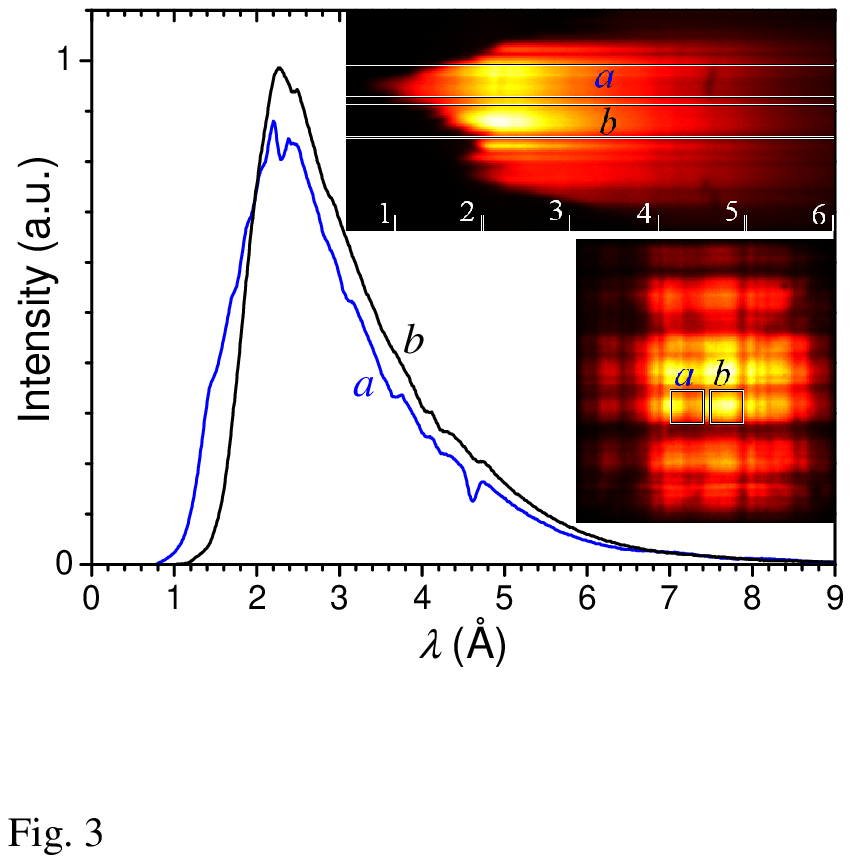}
\caption{Neutron spectra of regions of the monochromator crystals viewed by 
the pinhole-camera technique looking upstream from behind the GINA 
monochromator - through the graphite crystals (\textit{a}), and aside from them 
(\textit{b}). Dips at $~4.6~\mathrm{\mathring{A}}$ and $2.3~\mathrm{\mathring{A}}$ 
are the basic and the first harmonic 
deflections by the monochromator's graphite crystals. The difference 
between curves (\textit{a}) and (\textit{b}) are due to the neutron guide's Ni/Ti 
supermirror coating being viewed at different angles through the pinhole. 
Lower inset shows the detector image summed for the $4\div
5~\mathrm{\mathring{A}}$ interval. The upper inset is a horizontal divergence
vs.\ wavelength map summed for the height of the summation areas (squares with
sides of 29~mm) in the lower inset.}
\label{Fig3}
\end{figure}

The TOF neutron spectrum of the beam leaving this guide section was 
measured by the pinhole camera technique \cite{refe22} looking upstream 
from behind the GINA monochromator. The results are shown in Fig.~\ref{Fig3}. 
Spectrum (\textit{a}) and spectrum (\textit{b}) represent the directions 
of neutrons passing 
through the monochromator crystals and alongside of them, respectively. 
The parameters of the TOF experiment -- monochromator to pinhole 
distance $3100~\mathrm{mm}$, pinhole to detector distance 
(flight length) $3750~\mathrm{mm}$, pinhole diameter: $3~\mathrm{mm}$, 
chopper open time: $0.1~\mathrm{ms}$, bin time: $8~\mathrm{\mu s}$, 
detector gas absorption depth: $30~\mathrm{mm}$ -- 
resulted in a wavelength resolution \cite{Fuzi2006} of 
$\Delta \lambda/\lambda=0.01+0.12~\mathrm{\mathring{A}/\lambda}$, 
leading to $\Delta \lambda/\lambda=3.6\%$ at $4.6~\mathrm{\mathring{A}}$ 
and 6.2\% at $2.3~\mathrm{\mathring{A}}$, about 3 to 4 times worse 
than the actual monochromator resolution. Consequently, the dips appear 
wider and shallower in Fig.~\ref{Fig3} at $4.6~\mathrm{\mathring{A}}$ 
and $2.3~\mathrm{\mathring{A}}$ than they are in reality.
\footnote{During the preparation of the manuscript, the 
monochromator crystals have been exchanged and rearranged for optimum 
reflectivity, as well as further shielding was installed along the beam path, 
which resulted in a considerable increase of the reflected intensity 
to background ratio to $5\times10^{-6}$. These modifications, 
along with new filter/polarizer options will be reported upon elsewhere.}
\noindent
%\begin{figure}[p]

The TOF pinhole imaging measurement results in a 3-dimensional data 
set: horizontal and vertical pixel positions of a detector event in 
correlation with the pinhole distance give the corresponding divergence 
angles and time elapsed from chopper opening in correlation with flight 
distance give the neutron speed (energy, wavelength). A horizontal plane 
cut in this database (or a sum of several such cuts) yields the local 
(height averaged) intensity distribution as a function of horizontal 
divergence and wavelength (similar to Fig.~1 in Ref.~\onlinecite{refe22}, 
explained in more detail in Ref.~\onlinecite{fuzikonyv}).

\begin{figure}[t]
%\resizebox{.6\columnwidth}{!}{\includegraphics{Fig4.eps}}
\resizebox{4.0cm}{!}{\includegraphics[clip=true]{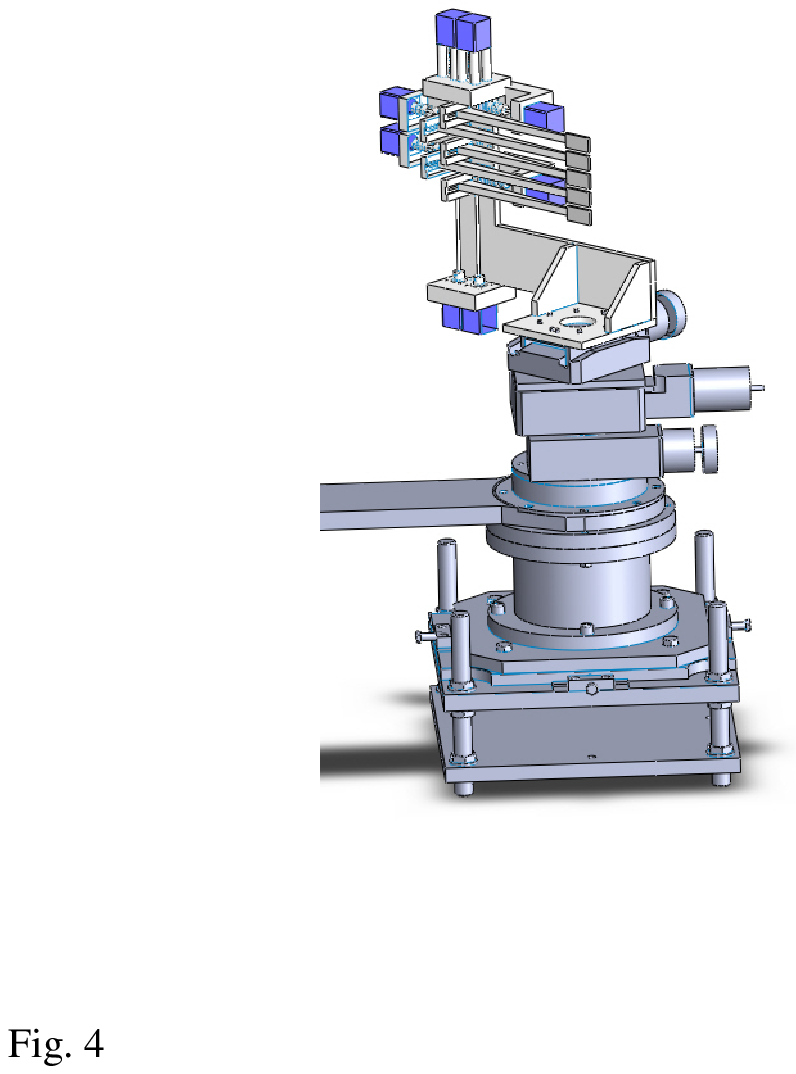}}
%\includegraphicss {Fig4.eps}
 \caption{The monochromator assembly of the GINA reflectometer. Besides
rotation, tilt and translation of the central graphite crystal (i.e.\ the full
monochromator), the top and bottom pairs of graphite crystals can be
individually rotated and tilted relative to it.}
 \label{Fig4}
\end{figure}

The lower inset of Fig.~\ref{Fig3} displays the detector image summed up 
for the $4\div5~\mathrm{\mathring{A}}$ wavelength interval. The upper 
inset is a horizontal divergence vs.\ wavelength map summed up for the 
height of the summation areas (squares with sides of 29~mm) in the 
lower inset. The monochromator assembly is mounted on a turntable 
connected to the optical bench B1 and comprises 
five highly oriented pyrolytic graphite (HOPG) crystals of 
$20\times20\times2~\mathrm{mm}^{3}$, attached to thin horizontal Al 
alloy rods (Fig.~\ref{Fig4}) and are lined up around the vertical axis.
\noindent
%\begin{figure}[p]

The mounting rod of the central graphite crystal is directly attached to the
monochromator bench, which is alined with respect to the axis of the
monochromator turntable and can be rotated around its vertical axis. The rods
of the remaining four crystals (two above and two below) are attached to small
motorized 2-axis goniometers for horizontal alignment and vertical focusing.

Vertical focusing to the sample position resulted in doubling the intensity
reflected by a $2\times2~\mathrm{cm}^{2}$ sample at grazing incidence as
compared to the non-focused case of parallel graphite crystals.

The entire monochromator assembly can be tilted around a horizontal axis. When
the instrument is not in operation, the monochromator assembly can be moved
out of the beam using a linear $x-$stage. All eleven motions mentioned above
are motorized and remotely controllable.

For the purpose of normalizing the measured reflectivity to the incident
intensity, a low efficiency $^{3}\mathrm{He}/\mathrm{CF}_{4}$ beam monitor
detector 
\footnote{Canberra XERAM, type MNH 10/4.2 SCAL} 
%\cite{refe23} 
with an active area of $H\times W=100\times42~\mathrm{mm}^{2}$ is
mounted in the beam path downstream of the monochromator assembly. The
detecting efficiency of the intensity monitor is $\approx0.1\%$ for
$\lambda=4.6~\mathrm{\mathring{A}}$ neutrons.

\begin{figure}[b]
%\resizebox{.45\columnwidth}{!}{\includegraphics{Fig6.eps}}
\includegraphicss {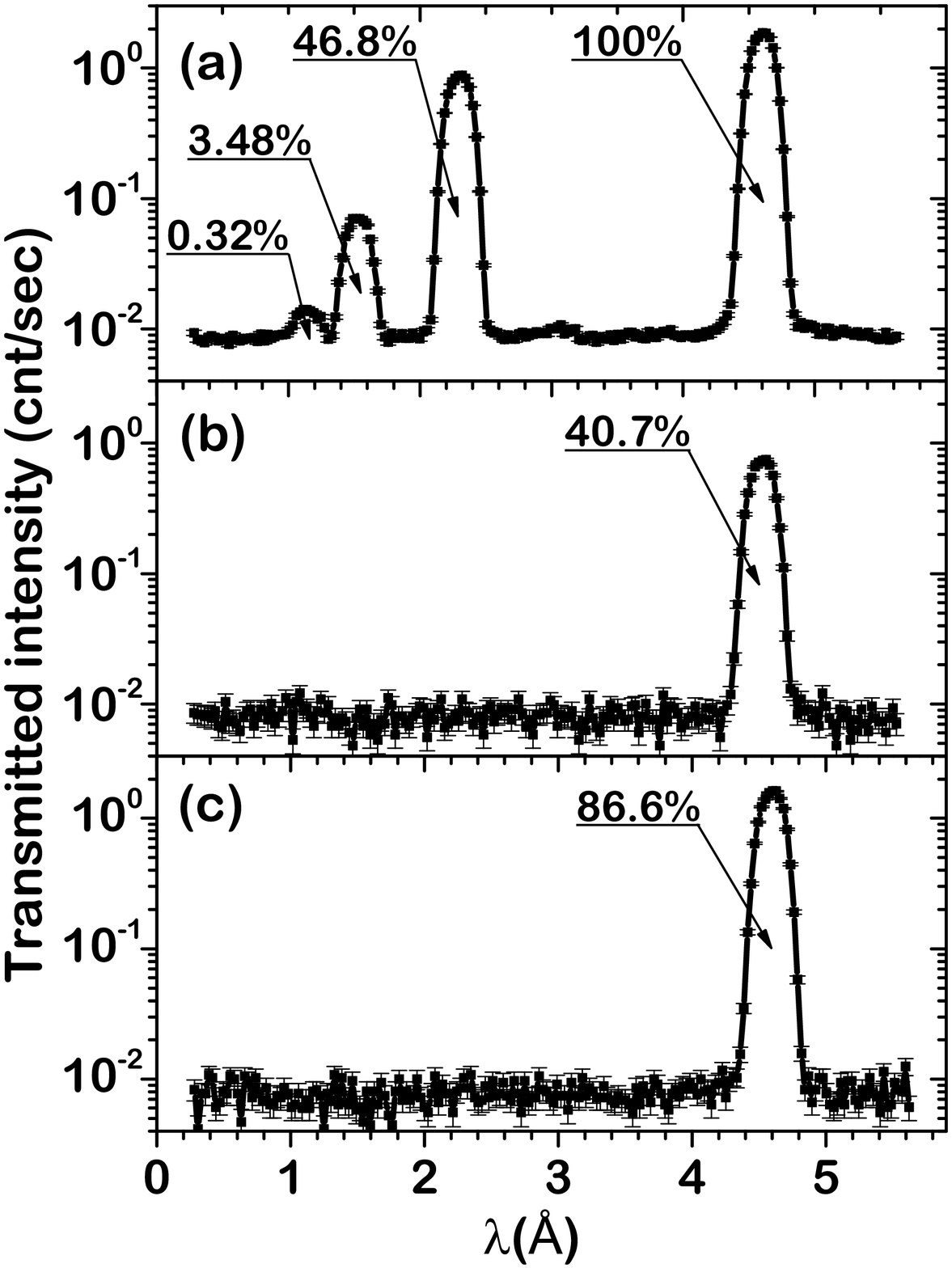}
\caption{Area-integrated TOF spectrum of the detector pictures in the 
parallel-aligned orientations of the HOPG crystals (not shown) without Be 
filter (a), and with Be filter at room temperature (b) and with 
Be filter cooled by liquid nitrogen (c). The normalized integrated 
intensity of the harmonics of increasing order 
are $66.4\%$, $31.1\%$, $2.3\%$ and $0.2\%$, respectively,\cite{Note3} i.e.\ the higher 
harmonics admixing is over $50\%$ at $\lambda=4.6~\mathrm{\mathring{A}}$.
The Be filter suppresses the higher harmonics at both room and liquid nitrogen 
temperature. The transmission of the Be filter is $40.7\%$ and $86.6\%$ for
$\lambda=4.6~\mathrm{\mathring{A}}$ at room temperature and in the cold state, 
respectively.}
\label{Fig6}
\end{figure}

Bragg reflections by the HOPG crystals contain higher harmonics according to
the energy distribution of the incident neutron beam. These
fractional-wavelength neutrons need to be filtered out from the beam incident
onto the sample. GINA is equipped with a Be filter with Be-slab size of
(height$\times$width$\times$depth) $72\times42\times150~\mathrm{mm}^{3}$ which
is cold-finger-cooled by liquid nitrogen during regular operation. The
transmission of the filter was measured using the TOF technique with a
wavelength resolution of 6\%. This experiment revealed that the filter has a
transmission of 40.7\% and 86.6\% for $\lambda=4.6~\mathrm{\mathring{A}}$,
without and with liquid nitrogen cooling, respectively, while suppressing all
higher harmonics at both temperatures~(Fig.~\ref{Fig6}).
\noindent
%\begin{figure}[p]

\section{SPIN POLARIZATION AND ANALYSIS}

\begin{figure}[b]
%\resizebox{.45\columnwidth}{!}{\includegraphics{Fig7.eps}}
\includegraphicss {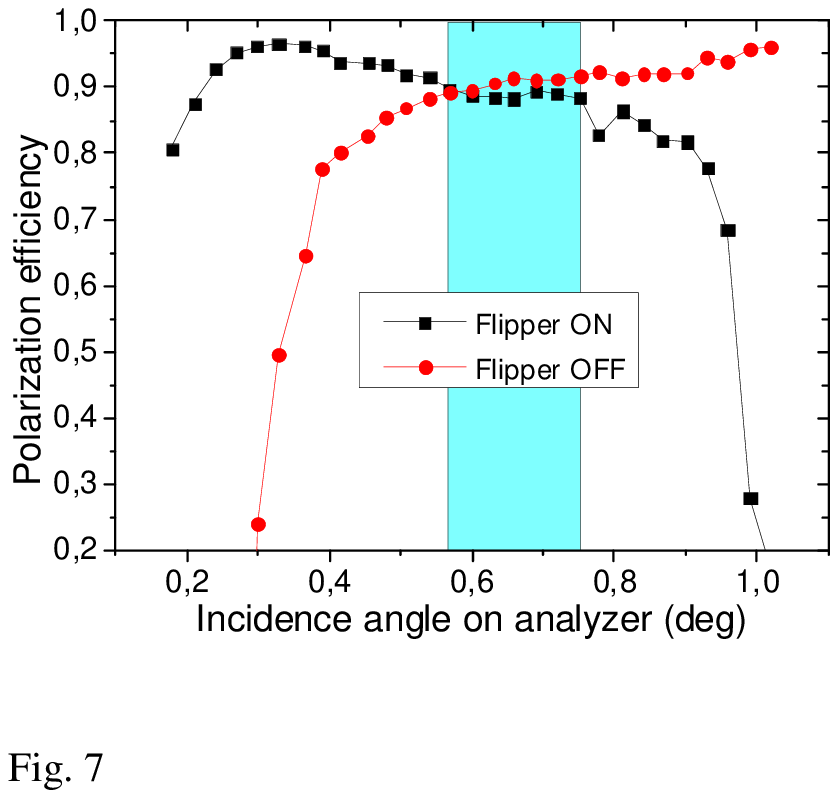}
 \caption{ Polarization efficiency of the GINA setup vs.\ the incidence angle on
the PSM of the analyzer P2, with SF1 ON and OFF, respectively, while SF2 was
kept OFF if both cases. The overall polarization efficiency is $\sim0.9$ in
the optimum angular range of operation which is marked by the rectangle.}
 \label{Fig7}
\end{figure}

In order to produce polarized neutrons an Fe-Co/Si magnetic supermirror 
polarizer (made by Neutron Optics, Berlin) is used. The vertically 
oriented Si wafers with $m=2.5$ supermirror coating are mounted onto a 
rotator and a translator for adjustment to optimum polarization efficiency 
(defined below). The supermirror is placed between yokes of a permanent magnet
construction with an in-plane vertical magnetic field of $30~\mathrm{mT}$. For
spin analysis of the reflected beam, a PSM analyzer is used in an identical
construction with the polarizer. By this setup, spin analysis of specular
scattering is easily possible. In the case of off-specular scattering detailed
studies are also possible but the experiments are rather time consuming. Both
supermirrors are used in transmission geometry. The corresponding overall 
polarization efficiency (of P1, SF1 and P2) is shown in Fig.~\ref{Fig7} 
as a function of the incident angle on the analyzer PSM.
\noindent
%\begin{figure}[p]

In neutron reflectometry the signal to background ratio is always a critical
parameter and has to be maximized. Suppression of scattering of neutrons by
the beam-line components is the key issue. Therefore, instead of using Mezei 
flippers of simpler construction (always wires in the beam) we decided to opt 
for RF spin flippers of own design and fabrication based on adiabatic 
fast passage (AFP). \cite{refe19,refe20} The flipper coil is placed in a transverse 
static magnetic field with longitudinal gradient, produced by two iron plates 
energized by Nd-Fe-B permanent magnet stacks upstream and 
shunted by soft iron rods downstream. The numerical simulation 
of the AFP process\cite{Fuzi2008calc} for design and virtual testing purposes 
has been performed based on the measured magnetic field 
distribution of the constructed frame. The RF coil for 
longitudinal RF field is part of a serial electric resonant 
circuit, with a sinusoidal current and bandwidth (full width at 
half maximum, FWHM) of $I_{\mathrm{eff}}=4~\mathrm{A}$ and 
$4.5~\mathrm{kHz}$ at the resonance frequency of $175~\mathrm{kHz}$. 
The RF current is provided by a remote 
controlled power supply. 
\footnote{Flipper power supply type ANSFR-83C (Promel Unlimited, Budapest, Hungary)}
%\cite{refe24} 
The parameters of the adiabatic RF 
spin flippers are summarized in Table~I. The flipper efficiency 
is better than 99\% at any wavelength above $2~\mathrm{\mathring{A}}$, 
as recently proven on flippers of similar design. \cite{saerbeck2012} 

\begin{figure}[t]
%\resizebox{.45\columnwidth}{!}{\includegraphics{Fig8.eps}}
\includegraphicss {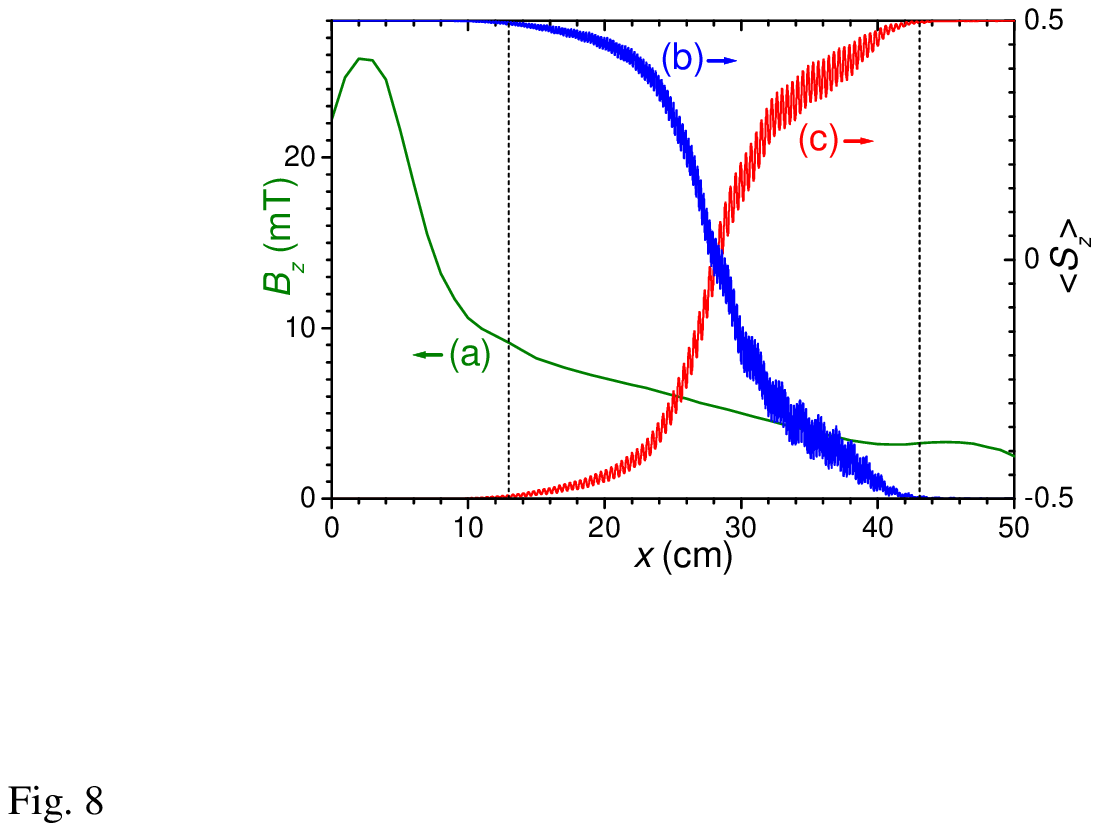}
 \caption{Simulation of operation of the adiabatic radio-frequency spin flipper of 
GINA, displaying the measured static vertical magnetic field, $B_{z}(x)$, 
(left scale, curve (a)) and the simulated evolution of the z-component of the neutron 
spin $\left<S_{z}\right>$ (right scale) along the coil axis, $x$ for spin-up and spin-down 
neutrons with wavelengths of  $5.5~\mathrm{\mathring{A}}$ (b) and $3.2~\mathrm{\mathring{A}}$ 
(c), respectively. Vertical short-dashed lines indicate the coil ends. The parameters 
of the simulation are summarized in Table~I.}
 \label{Fig8}
\end{figure}

The present flipper design is insensitive to the external field 
variations and/or influence of magnetic components along the beam path. It 
can be used in the entire wavelength range of GINA without further 
adjustment. The static field and the simulated projection of the magnetic 
moment of the neutron onto the direction of the field are shown 
in Fig.~\ref{Fig8} as a function of the neutron position along the beam 
path within the yokes of the flipper. For sake of clarity, the curves 
for two different wavelength values are plotted for spin-up and spin-down 
incoming spin states, respectively. 
Should the flipper terminate abruptly, i.e.\ at a given position $x$ both the 
static gradient field and the RF field become zero (a situation neither 
desired nor feasible), the neutrons would emerge with spin-state 
probability determined by the value of the respective curve at that 
position.
%{{{Therefore, instead of using Mezei
%flippers of simpler construction (always wires in the beam) we decided to opt 
%for RF spin flippers based on adiabatic fast passage. \cite{refe19,refe20} 
%The flipper coil is placed in a
%transverse static magnetic field with longitudinal gradient, produced by two
%iron plates energized by Nd-Fe-B permanent magnet stacks upstream and shunted
%by soft iron rods downstream. The RF coil for longitudinal RF field is part of
%a serial electric resonant circuit, with a sinusoidal current and bandwidth
%(full width at half maximum, FWHM) of $I_{\mathrm{eff}}=4~\mathrm{A}$ 
%and $4.5~\mathrm{kHz}$ at the resonance frequency of $175~\mathrm{kHz}$. The RF
%current is provided by a remote controlled power supply. 
%%\footnote{Flipper power supply type ANSFR-83C (Promel Unlimited, Budapest, Hungary)}
%%\cite{refe24} 
%The parameters of the adiabatic RF spin flippers are summarized in Table 1. The flipper
%efficiency is better than 99\% at any wavelength above $2~\mathrm{\mathring{A}}$, 
%as recently proven on flippers of similar design.\cite{saerbeck2012} 
%The present flipper design is insensitive to the external field
%variations and/or influence of magnetic components along the beam path. It can
%be used in the entire wavelength range of GINA without further adjustment. The
%static field and the simulated projection \cite{refe34} of the neutrons' 
%magnetic moment onto the direction of the field are shown in Fig.~\ref{Fig8} 
%as a function of the neutron position along the beam path.}}}
\noindent
%\begin{figure}[p]

For neutron spin analysis an identical set of adiabatic RF spin flipper and
PSM is placed downstream to the sample. The overall polarization efficiency 
of the setup was determined by measuring the reflected and transmitted 
intensity on the analyzer P2 (placed in the sample position) with and 
without activating the spin flipper SF1, while spin flipper SF2 was kept OFF.
The polarization efficiency was calculated by the formula%
\begin{equation}
P=\frac{I^{+}-I^{-}}{I^{+}+I^{-}} \label{Eq1}%
\end{equation}
where $I^{+}$ and $I^{-}$ are the corresponding reflected and transmitted
intensities by the analyzer P2. This yields an overall efficiency of 0.9 for
P1, SF1 and P2.

\begin{table}[ht]
\caption{Parameters of the adiabatic RF spin flippers}
\begin{tabular}
[c]{lr}\hline \hline
\textbf{Parameter} & \textbf{Range}\\ \hline
beam height & 60~mm\\
wavelength range & $\left(  3.1\div5.8\right)  ~\mathrm{\mathring{A}}$\\
frequency & 175~kHz\\
coil length / coil diameter & 300~mm~/~60~mm\\
number of turns & 100\\
inductance / capacitance & 95~$\mu$H~/~8.8~nF\\
voltage / current (effective) & 40~V~/~4~A\\
magnet block (Nd-Fe-B) $2\times$ & $60\times50\times20$~mm$^{3}$\\
yoke $\left(  L\times W\times H\right)  $ $2\times$ & $500\times100\times
12$~mm$^{3}$\\
magnetic field in the centre & 5.6~mT\\
longitudinal gradient  & $\left(  0.2\div0.4\right)  $~mT/cm\\ \hline \hline
\end{tabular}
\end{table}

\section{SAMPLE POSITIONING}

The flat sample is mounted on an adjustable vertical flat surface attached to
the top cradle of the central goniometer tower. The sample mounting depends on
the sample environment. For room temperature reflectivity measurements the
flat surface has a small bore through which the sample is sucked to the
vertical surface and held in position during the experiment by a small vacuum
pump. In such a way undesired scattering by the fixing elements is minimized.
Symmetrical sample positioning is ensured by using two cradles and two
perpendicular linear stages. The cradles and translators position the sample
in the vertical plane and set the sample surface orientation. The $\theta$ and
$2\theta$ angles are encoder controlled for increased precision. Fine
positioning of the beam is maintained by several slits with cadmium blades.
The slits can be opened in the range of 0 to $10~\mathrm{mm}$ with a precision
better than 0.2~mm. One slit (S1) is placed downstream of the Be filter and
one (S2) downstream of flipper SF1, just upstream of the sample. The slit S1
defines the beam on the polarizer mirror to decrease the beam divergence thus
to increase the polarization ratio. Slit S2 decreases the beam divergence on
the sample and absorbs the neutrons which might not reach the sample or
scattered off by the polarizer. With these optical elements the setup can
achieve a relative $Q$-resolution of 10\% to 2\% for the available $Q$-range
of 0.005 to 0.25$~\mathrm{\mathring{A}}^{-1}$.

\section{SAMPLE ENVIRONMENT}

GINA is primarily dedicated to reflectometry of magnetic heterostructures. For
studies of magnetism, vital environmental parameters are (low) temperature and
(occasionally high) applied magnetic field. A closed cycle $^{4}\mathrm{He}$
cryostat (comprising a cold finger setup 
\footnote{The cold finger cryostat 
is of Edwards CoolStar Coldhead 2-9 type, driven by an Edwards Cryodrive 3 
closed cycle He compressor} 
%\cite{refe25} 
and evacuated by a small
turbomolecular pump) can be mounted on the central goniometer tower of GINA
with or without the electromagnet. The sample temperature can be varied in the
nominal 9 to $300~\mathrm{K}$ range. 
\footnote{Temperature control is performed 
by a Model 336 Lakeshore controller using standard Pt-100 and carbon glass sensors} 
%\cite{refe26} 
The GINA beam line is equipped with an
air-cooled electromagnet which produces magnetic fields up to $0.55~\mathrm{T}%
$ for the pole distance of $40~\mathrm{mm}$ that accommodates the
1.5\textquotedblright \ diameter Al cap of the cryostat. A water-cooled air
core coil pair provides smaller magnetic fields up to approx. $35~\mathrm{mT}$.

\section{NEUTRON DETECTION AND BACKGROUND REDUCTION}

For detecting the neutrons, a commercial position sensitive multi-wire 
proportional chamber is used (made by Mirrotron Ltd., Budapest).
It is of $200\times200~\mathrm{mm}^{2}$ active area and is filled with 
$^{3}\mathrm{He}/\mathrm{CF}_{4}$  mixture of $3/2$~bar 
partial pressures and has a nominal spatial resolution 2~mm, and 
Al-window absorption and gas absorption efficiency 
of 8\% and 94\%, respectively at $4.6~\mathrm{\mathring{A}}$.
In order to suppress the background, the detector is encased in
a polyethylene shielding of 30~mm thickness containing $20~\mathrm{wt}\%$ natural B$_{2}%
$O$_{3}$. The two-dimensional spatial detection is managed by two delay lines
and the positions are determined by a DASY TDC module (produced by ESRF,
Grenoble) installed in a slot of the detector PC which is dedicated
exclusively to the detector data acquisition and mounted on the $2\theta$-arm
of the reflectometer. When detecting specular scattering, two slits (S3 and
S4) are placed in front of the detector window to discard undesired radiation.
If no spin analysis is required, for further background suppression, an
evacuated flight tube is mounted along the entire length of the $2\theta$
detector arm. Mounting the spin analyzer and flipper in a vacuum vessel is a
plan for the future. The basic operation parameters of the GINA reflectometer 
are summarized in Table~II.

\section{INSTRUMENT CONTROL}

The GINA hardware and the control software are rather flexible and are
designed for maximum remote controllability. In its full configuration, GINA
comprises more than 30 stepping motors. 
\footnote{The stepping motors are controlled by MCU-2FX and StepPack controllers 
from Advanced Control System Corporation, USA or SixPack 2 integrated 
controllers produced by TRINAMIC Motion Control GmbH, Germany.}
%\cite{refe27} 
Certain critical motions, such as
$\theta-$ and $2\theta-$ angles and precision slit positions are absolute or
relative encoder-controlled. A custom made unit built around a USB
multi-function data acquisition module 
\footnote{USB data acquisition module of type DT 9802, produced by Data 
Translation, USA}
%\cite{refe28} 
controls the air compressor, the air
pads, the liquid nitrogen level and temperature in the Be-filter, the beam
shutter and its control lights, the beam intensity monitor and the various
modular DC power supplies (the latter ones to energize electromagnets, and
various coils in the setup including optional Mezei flippers). The high
voltage power supplies of the detector and that of the beam monitor, the
linear amplifiers, the discriminators and the ratemeters are realized in
modules of NIM standard. The control PC directly communicates with the
detector PC via ethernet and with the indexer modules of the motion control
units as well as with the temperature controller via RS232 lines. All listed
components are controlled by the GINASoft control software written in LabView
V10.0 for MS Windows. The user interface of the program allows for various
alignment and scan modes as well as changing polarization and sample
environment (flipper current and frequency, temperature, magnet current, etc.)
remotely. For increased user comfort the command format of the user interface
is user-configurable, and includes a command format that resembles to that of
SPEC. 
\footnote{SPEC by Certified Scientific Software, \protect\url{http://www.certif.com}}
%\cite{refe29} 
2D detector pictures and reduced reflectivity data can be efficiently
viewed and manipulated during data acquisition. Collected and manipulated data
as well as extended log information (including graphics) are saved in a
clearly structured database format. Human control is facilitated by a web
camera which is installed in the control PC. Using remote desktop option, most
operations can be performed remotely via internet from outside the
experimental hall or even from a distant continent.
\begin{table}[t]
\caption{Operation parameters of the GINA neutron reflectometer 
at the Budapest Neutron Centre}
\begin{tabular}
[c]{p{4.5cm} r}\hline \hline
\textbf{Parameter} & \textbf{Range}\\ \hline
wavelength & $\left(  3.2\div5.7\right)  ~\mathrm{\mathring{A}}$ in five
steps\\
present wavelength & $4.6~\mathrm{\mathring{A}}$\\
max. scattering angle $\left(  \theta \right)  $ & $\geq35%
%TCIMACRO{\U{b0}}%
%BeginExpansion
{{}^\circ}%
%EndExpansion
$\\
angular resolution $\left(  \Delta \theta \right)  $ & $0.003%
%TCIMACRO{\U{b0}}%
%BeginExpansion
{{}^\circ}%
%EndExpansion
$\\
$\Delta \lambda/\lambda$ & $\sim1\%$\\
\hspace{0 pt}background level at the 2D detector & $0.01$~cps~cm$^{-2}$\\
detector & 2D PSD $200\times200~\mathrm{mm}^{2}$\\
detector resolution & $2\times2$ mm$^{2}$\\
\hspace{0 pt}neutron flux at the monochromator exit & $4\times10^{5}~\mathrm{n}%
\times \mathrm{cm}^{-2}\times \mathrm{s}^{-1}$\\
background reflectivity\cite{Note3} & $<3\times10^{-5}$\\ 
(unpolarized, 4" Si) \\ 
\hspace{0 pt}overall polarization efficiency\cite{Note3} of polarizer and analyser SM & $0.9$%
\\ \hline \hline
\end{tabular}
\end{table}

\section{DATA HANDLING AND EVALUATION}

Users of the GINA reflectometer are offered the data handling and evaluation
software FitSuite, \cite{Sajti2009,*refe30a} a thoroughly documented program with a detailed project
home page, written for Windows \textit{and} Linux, which is presently suitable
for evaluating data of 14 experimental methods, including specular polarized
neutron reflectometry (with e.g. model for diffusion, also in isotope-periodic
multilayers) and off-specular (diffuse) polarized neutron reflectometry (in
the distorted-wave Born approximation) as well as specular x-ray
reflectometry. Specular and off-specular reflectivities are calculated
using the matrix approach  Ref. \onlinecite{refe3} and the Distorted Wave Born 
Approximation given in Ref. \onlinecite{Deak2007}. FitSuite \cite{Sajti2009,*refe30a} 
handles the corresponding theories and sample structures consistently 
in a common structure and allows for parameter restrictions, correlations 
and simultaneous simulation and fit of models to the experimental data.

\section{EXAMPLE REFLECTOGRAMS}

\begin{figure}[t]
%\resizebox{.45\columnwidth}{!}{\includegraphics{Fig9.eps}}
\includegraphicss {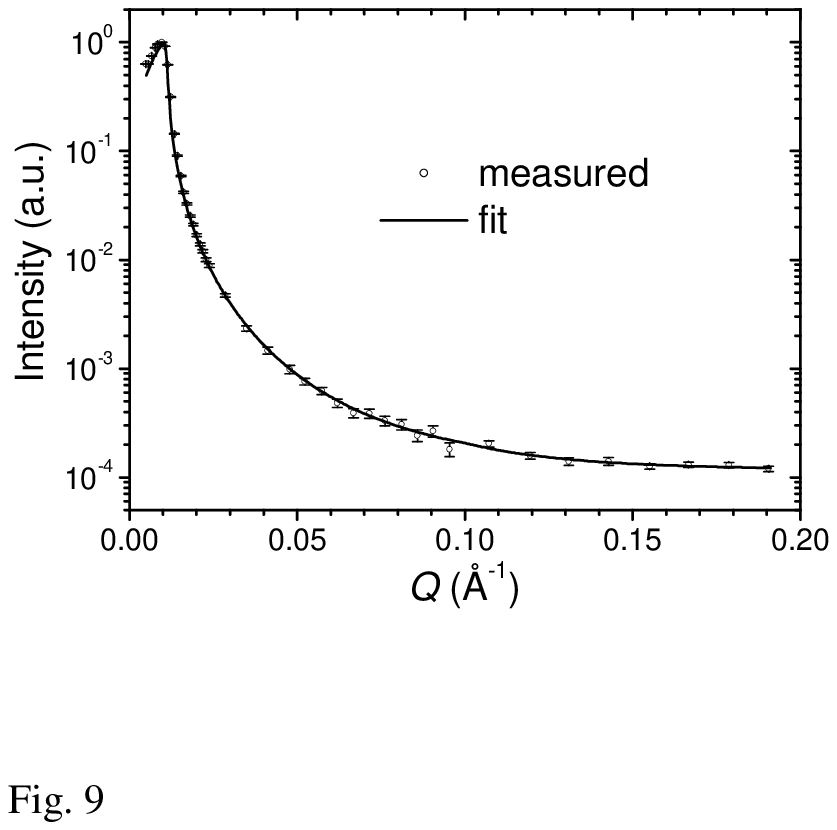}
 \caption{The specular reflectivity curve of a four-inch Si wafer measured at
the GINA reflectometer in non-polarized mode. The fitted background
reflectivity is $3\times10^{-5}$.}
 \label{Fig9}
\end{figure}

Two example reflectograms are chosen to highlight the present performance of
the GINA setup. The first one is the specular reflectivity of a 4-inch Si
wafer measured in non-polarized mode. The data shown in Fig.~\ref{Fig9} were
collected over a time of 22 hours. The fit to the experimental points gave an
SLD value of $(3.03\pm0.03)\times10^{-6}~\mathrm{\mathring{A}}^{-2}$ for the
native surface oxid layer of 20$~\mathrm{\mathring{A}}$ thickness and
roughnesses of 4$~\mathrm{\mathring{A}}$ for both the silicon and the native
oxide surfaces, respectively. The SLD for the silicon was kept constant to the
nominal value of $2.07\times10^{-6}~\mathrm{\mathring{A}}^{-2}$. The obtained
SLD value for the oxide layer is slightly smaller than the SLD value of
$3.47\times10^{-6}~\mathrm{\mathring{A}}^{-2}$ for amorphous bulk SiO$_{2}$.
This deviation may result from the lower density of the thin oxide layer as
compared to that of the bulk. The fit, full line in Fig.~\ref{Fig9}, provides a
background reflectivity of $3\times10^{-5}$.
\noindent
%\begin{figure}[p]

\begin{figure}[b]
%\resizebox{.45\columnwidth}{!}{\includegraphics{Fig10.eps}}
\includegraphicss {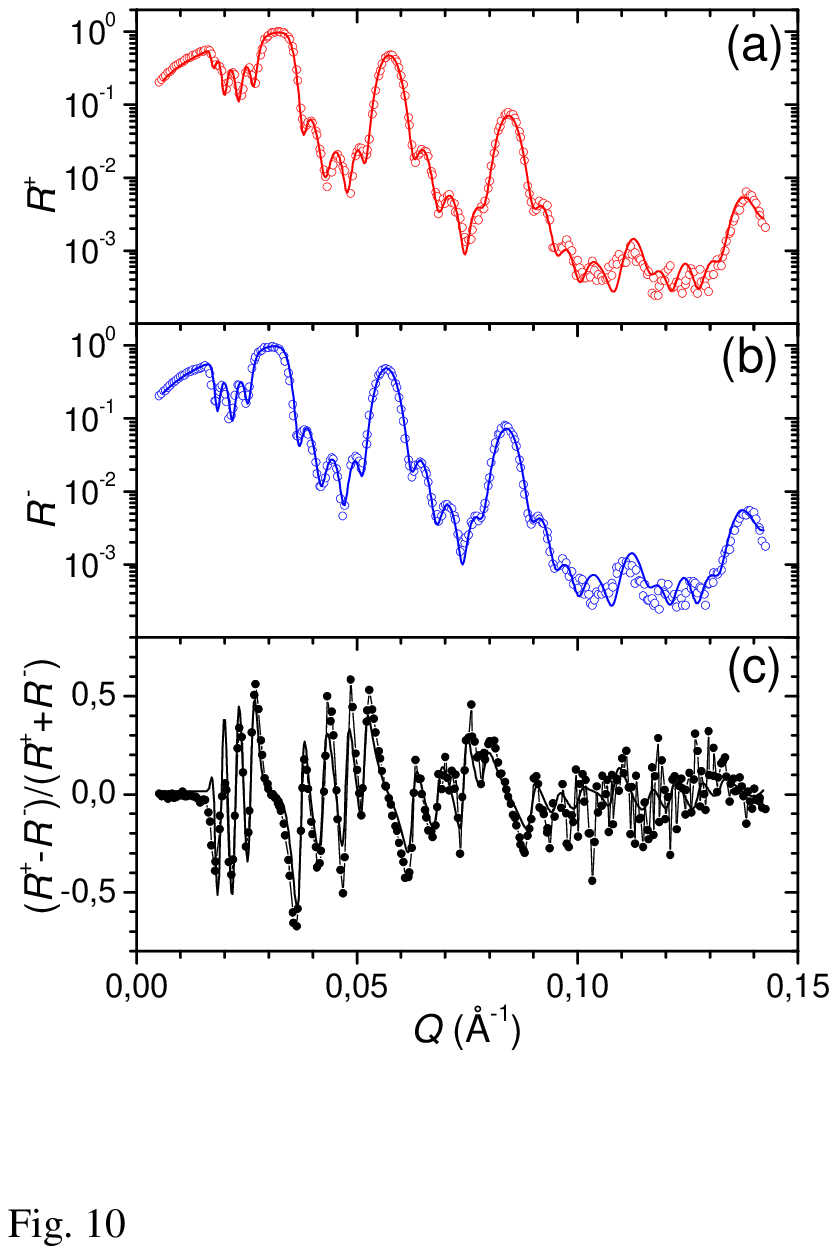}
 \caption{ Measured spin-up $R^{+}$, and spin-down $R^{-}$ reflectivities and
the calculated $\left(  R^{+}-R^{-}\right)  /\left(  R^{+}+R^{-}\right)  $
spin asymmetry of the isotope-periodic multilayer $\mathrm{MgO}\left(
001\right)  /[^{62}\mathrm{Ni}(5~\mathrm{nm})/^{\mathrm{nat}}\mathrm{Ni}%
(15~\mathrm{nm})]_{5}$ measured on the GINA reflectometer in polarized mode
without polarization analysis.}
 \label{Fig10}
\end{figure}

The second example is a magnetized Ni film. Polarized neutron reflectometry
provides a means to simultaneously determine the atomic layer and
magnetization profile in a multilayer. Deviation of both atomic and magnetic
momentum density of Ni films from the bulk values was a subject of earlier
neutron reflectometric studies. Singh and Basu studied a Ni film of
1500$~\mathrm{\mathring{A}}$ on glass substrate (a neutron mirror) by
unpolarized and polarized neutron reflectometry. \cite{refe31,refe32} Their analysis of the data
provided a nearly 50\% decrease of the atomic and a similar extent of decrease
in the magnetic momentum density in a surface layer of 235$~\mathrm{\mathring
{A}}$ and an about 10\% decrease of both quantities in the rest of the film
without any chemical change. Therefore we decided to study such effects on a
carefully prepared Ni film. In order to exclude undesired surface effects, an
isotopic multilayer Ni film of nominal layer structure $\mathrm{MgO/[}%
^{62}\mathrm{Ni(5nm)/}^{\mathrm{nat}}\mathrm{Ni(15nm)}]_{5}$ was prepared by
molecular beam epitaxy in the Wigner Research Centre for Physics, Budapest
depositing Ni of natural isotopic abundance by electron beam evaporation and
96.2\% enriched $^{62}$Ni from a Knudsen cell onto $\left(  100\right)  $ MgO
of $20\times20\times2~\mathrm{mm}^{3}$ size. Reflection high energy electron
deposition images taken during deposition revealed both epitaxial and
polycrystalline regions in the multilayer. Due to the large negative SLD of
$^{62}$Ni, it provides a sensitive probe of any structural change in the
multilayer. Specular reflectivity of the sample was measured on the GINA
reflectometer at room temperature in polarized mode without polarization
analysis. Prior to the experiment the sample was magnetized from the virgin
state by 50~mT in-plane external magnetic field parallel to the guide field.
The total data collection time was 56 hours. In Figs.~\ref{Fig10}a--\ref{Fig10}b
the reflectivities $R^{+}$ and $R^{-}$ are shown together with the curves of
the simultaneous fit by FitSuite \cite{Sajti2009,*refe30a} (latter in full lines). For clarity, the
measured and fitted spin asymmetry $\left(  R^{+}-R^{-}\right)  /\left(
R^{+}+R^{-}\right)  $ curves are displayed separately in Fig.~\ref{Fig10}c. 
  \noindent
%\begin{figure}[p]
The
simultaneous fit was constrained to a periodic layer structure and yielded
layer thicknesses of $(175\pm5)~\mathrm{\mathring{A}}$ and $(53.5\pm
5)~\mathrm{\mathring{A}}$ for $^{\mathrm{nat}}$Ni and $^{62}$Ni, respectively
and common rms interface roughness of $(5\pm2)~\mathrm{\mathring{A}}$. The
errors are $1\sigma$ statistical errors as obtained by the least-squares fit.
The fitted scattering length densities were: $\mathrm{SLD(}^{\mathrm{nat}%
}\mathrm{Ni)}=(9.13\pm0.5)\times10^{-6}~\mathrm{\mathring{A}}^{-2}$;
$\mathrm{SLD(}^{62}\mathrm{Ni)}=(-7.0\pm1)\times10^{-6}~\mathrm{\mathring{A}%
}^{-2}$ showing only minor deviations from their known bulk SLD values,
consequently as well as from the bulk Ni atomic density. The magnetization in
the $^{\mathrm{nat}}$Ni and $^{62}$Ni layers were kept identical in the fit
which provided $(0.44\pm0.12)~\mathrm{T}$. This value amounts only to 66\% of
the known room temperature saturation magnetization MS of the bulk Ni
($55~\mathrm{emu/g}$, \cite{refe33} corresponding to $0.67~\mathrm{T}$). This may be either
due to a decrease of the Ni magnetic moment in the layers or an incomplete
magnetic saturation of the sample. The reflectivity measurement was performed
in 50~mT external magnetic field. This in-plane field strength was considered
sufficient for saturation, since the bulk Ni saturates at 300~Oe even along
the $\left(  100\right)  $ direction of the hard magnetization. \cite{refe33}

\begin{figure}[ht]
%\resizebox{.45\columnwidth}{!}{\includegraphics{Fig11.eps}}
\includegraphicss {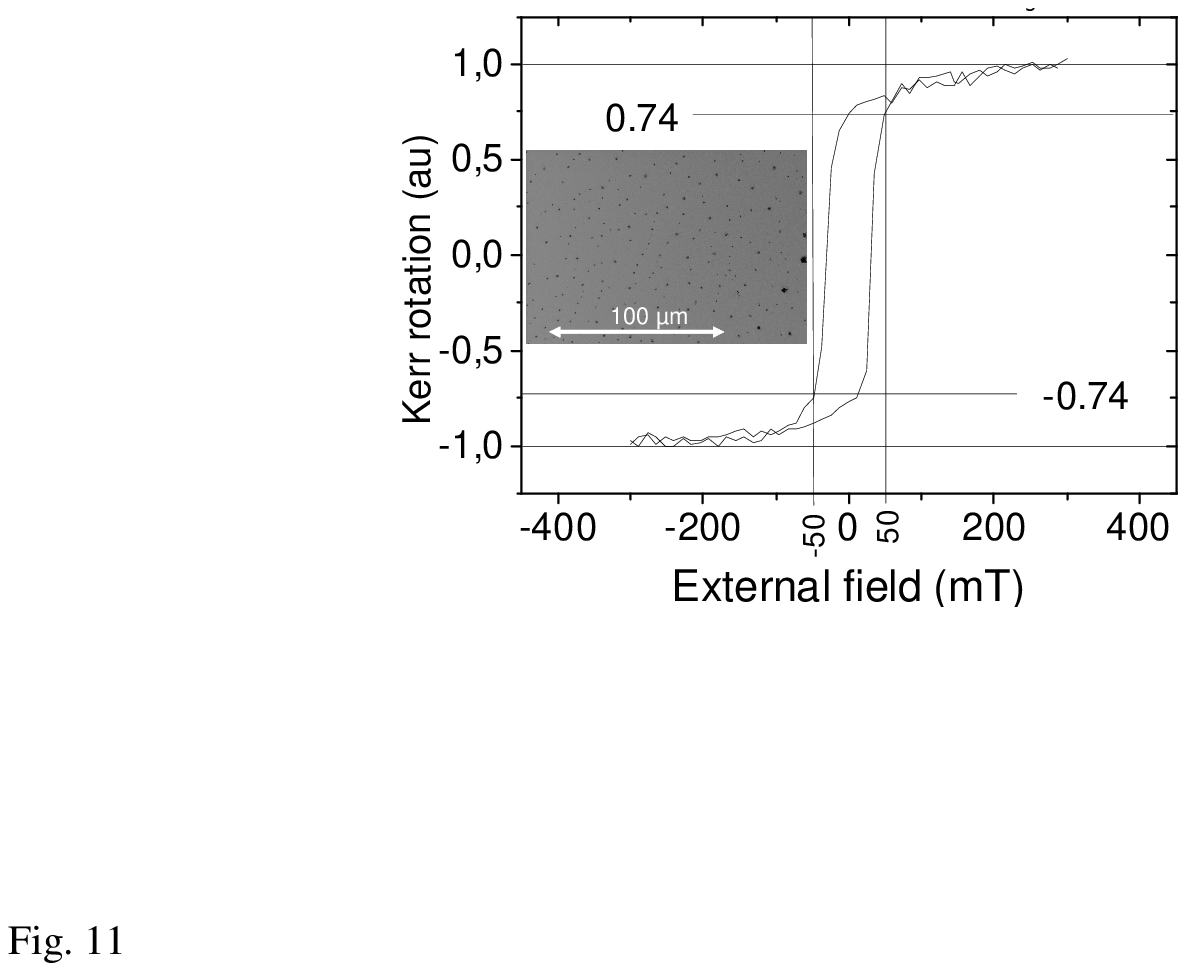}
 \caption{Magnetooptical Kerr loop of the Ni isotope multilayer sample
showing an incomplete magnetic saturation of the layer at 50~mT, the external
field used in the reflectivity measurement. Inset: Scanning electron
microscope picture of a $200\times130$~%
%TCIMACRO{\U{3bc}}%
%BeginExpansion
$\mu$%
%EndExpansion
m surface region of the sample. Tubular discontinuities amount up to about
$1\%$ of the film volume.}
 \label{Fig11}
\end{figure}

However, the magnetooptical Kerr-loop (Fig.~\ref{Fig11}) taken subsequent to the
reflectivity measurement revealed a partial (74\%) saturation of the Ni film
in 50~mT. The somewhat lower value of 66\% provided by the reflectivity fit
can be explained by a lower magnetization of the near-edge regions of the
sample, contributing to the reflectivity but not to the MOKE signal. Scanning
electron microscopic pictures of the sample surface (inset in Fig.~\ref{Fig11})
reveals sparse plane-perpendicular tubular discontinuities of the multilayer.
Image analysis provides 1.37\% surface coverage of the tube openings.
Therefore the total volume of the tubes and their effect on the average film
density is negligible in the present sample.
  \noindent
%\begin{figure}[p]

\section{SUMMARY}

We have shown that GINA, the newly installed dance-floor-type constant energy
angle-dispersive neutron reflectometer at the Budapest Neutron Centre is a
versatile instrument in both polarized and unpolarized modes of operation.
Examples were given for both modes of operation. The sample orientation of the
reflectometer is vertical. The available sample environment facilities are:
closed cycle cryostat optionally combined with external magnetic field by an
iron core electromagnet up to 0.55~T or air core coils for 35~mT in various
directions, spin polarization and polarization analysis by single polarizing
supermirrors. Detection of specular and diffuse scattering is facilitated by a
two-dimensional position-sensitive detector. All components of the instrument
are controlled by a program written in LABVIEW. The program allows for
alignment and scan modes as well as changing polarization and sample
environment parameters remotely. Reflectivities above five orders of magnitude
have been measured with further improvements underway. Further developments
including an environmental cell for biomimetic membrane studies, an
electromagnet with higher fields and orientation versatility and a supermirror
fan analyzer and further background suppression elements are also planned. The
GINA reflectometer is open for Hungarian and international users throughout
the year.
\footnote{Information concerning proposal submissions can be found at
\protect\url{www.bnc.hu}}
%\cite{refe35}

\begin{acknowledgments}
The GINA team is grateful to Prof. H. Dosch, former director of
Max-Planck-Institut f\"{u}r Metallforschung for his continued interest in the
GINA project and for the transfer of a number of components of EVA, a former
neutron reflectometer operated by the Max-Planck-Institut f\"{u}r
Metallforschung, Stuttgart at the Institut Laue-Langevin, Grenoble, France.
Important support obtained from the members of the Stuttgart neutron group,
namely from A. Vorobiev and P. Falus, (Grenoble) and A. R\"{u}hm and J. Franke
(Garching) is deeply appreciated. Helpful advises by Yu. V. Nikitenko (Frank
Laboratory of Neutron Physics, JINR, Dubna, Russia) at an early stage of the
GINA project are gratefully acknowledged. We are also grateful to T. Keller,
(Max-Planck-Institut f\"{u}r Festk\"{o}rperforschung, Stuttgart) for his
valuable support in the later stage of the construction work. Authors are
thankful for the electronic and mechanical design and construction work on the
GINA components to the colleagues at the Wigner Research Centre for Physics), 
Budapest, (formerly KFKI Research Institute for Particle and Nuclear Physics)
in particular to P.~Ruszny\'{a}k, J.~Gigler, G.~Gy.~Kert\'{e}sz and L.~Jakab, 
as well as the mechanical workshop team led by F.~Bodai. The support of the
management and staff of the reactor of the Budapest Neutron Centre and
preparation of the $^{62}$Ni/ $^{\mathrm{nat}}$Ni multilayer sample by F.
Tanczik\'{o} are gratefully acknowledged. This work was partially supported by
the National Office for Research and Technology of Hungary and the Hungarian
National Science Fund (OTKA) under contracts NAP-VENEUS'05 and K~62272,
respectively. Mobility support for A.V. Petrenko by the bilateral project
between JINR (Dubna) and the Hungarian Academy of Sciences is gratefully appreciated.
\end{acknowledgments}

%\bibliographystyle{apsrev4-1}
%\bibliography{acompat,gina-120607.bib}
\newif\ifabfull\abfulltrue

\end{document}